\documentclass[a4paper]{jpconf}
\usepackage{graphicx}
\begin{document}
\title{High-frequency VLBI observations of Sgr\,A* during a multi-frequency campaign in May 2007}

\author{
R-S Lu$^{1,2}$, T P Krichbaum$^1$, A Eckart$^2$, S K\"onig$^{2,1}$, D Kunneriath$^{2,1}$, G Witzel$^2$, A Witzel$^1$ and J A Zensus$^1$ 
}

\address{$^1$ Max-Planck-Institut f\"ur Radioastronomie, Auf dem H\"ugel 69, D-53121 Bonn, Germany}
\address{$^2$ University of Cologne, I. Physikalisches Institut, Z\"ulpicher Str. 77, D-50937 K\"oln, Germany}

\ead{rslu@mpifr-bonn.mpg.de}

\begin{abstract}
In May 2007 the compact radio source Sgr\,A* was observed in a global multi-frequency monitoring campaign, from radio to X-ray bands. Here we present and discuss first and preliminary results from polarization sensitive VLBA observations, which took place during May 14-25, 2007. Here, Sgr\,A* was observed in dual polarization on 10 consecutive days at 22, 43, and 86\,GHz. We describe the VLBI experiments, our data analysis, monitoring program and show preliminary images obtained at the various frequencies. We discuss the data with special regard also to the short term  variability.

\end{abstract}

\section{Introduction}
There is overwhelming evidence that  Sagittarius\,A* (Sgr\,A*), the extremely compact radio source at the center of our Galaxy, is associated with a super-massive black hole.  It shows sudden bursts of radiation (called flares) a few times per day, which are thought to be related to the guzzling of gas and dust from its environment and which can be detected in X-ray and near infra-red (NIR) regime (e.g., \cite{eckart08} and references therein). To search for correlated short term variability and to investigate the flares in as many wavebands as possible, a world wide radio--sub-mm--NIR--X-ray observing campaign on Sgr\,A* has been carried out in May 2007, making use of many telescopes including in the radio bands the VLBA (Very Long Baseline Array), GMVA (The Global Millimeter VLBI Array), CARMA (Combined Array for Research in Millimeter-Wave Astronomy), ATCA (Australia Telescope Compact Array), IRAM (Institut de Radio Astronomie Millimetrique), and Effelsberg.

The VLBA observed Sgr\,A* at 22, 43, and 86\,GHz on 10 consecutive days. These data are used to measure the source size and to search for possible deviations from point-like (or symmetric) structure,  and variations of the source on time scales of $\sim$ 1 week. Since the intrinsic source size and shape is masked by the interstellar scattering at centimeter wavelengths, the combination of our 3 observing frequencies is of great importance for the size determination and is crucial for discriminating interstellar broadening from the source intrinsic structure of Sgr\,A* by the known scattering law (e.g., \cite{bower06}).

Here we show and discuss preliminary images obtained with the VLBA at the 3 frequencies and focus our discussion on the results mainly obtained at 43\,GHz. Other results regarding to the total flux density measurements in this global campaign are presented in other papers (see Kunneriath \textit{et al}, Eckart \textit{et al}, this conference).

\section{Observations and data analysis}
\subsection{Observations}
The observations were performed with the VLBA on 10 consecutive days during May 15 - 24, 2007 at 22, 43, and 86\,GHz, respectively. Each station recorded dual polarization with a sample rate of 512\,Mbps (8 \,intermediate frequency (IF) channels, 16\,MHz per IF, and 2\,bits per sample). In the frequency switching more VLBI time was spent at the shorter wavelength. Bright quasars (NRAO\,530, PKS\,1749+096, 3C\,279) served as amplitude calibrators and fringe tracers. The SiO maser in VX\,Sgr (transitions $v$=1, $J$=1,0 and $J$=2--1) was observed in interleaved short VLBI scans at 43 and 86\,GHz in order to complement the regular system temperature measurements for an improved amplitude calibration. The data were correlated at the VLBA AOC correlator in Socorro, NM, USA with 1\,s integration time.

\subsection{Data analysis}
The data were analyzed in AIPS using the standard algorithms including phase and delay calibration and fringe fitting. The amplitude calibration was performed with the measurements of the antenna gain and system temperatures for each station. Atmospheric opacity corrections were applied using the AIPS task ``APCAL''. Images of Sgr\,A* were finally produced using the standard hybrid mapping methods at all three frequencies. In Figure~1, we show as example the first epoch resulting hybrid images at each frequency. Table~1 parameters these images. During the imaging process, we verified correctness of the station gain calibration by comparing the amplitude calibration of Sgr\,A* with those of NRAO\,530, VX\,Sgr, and PKS\,1749+096.

\section{High frequency images of Sgr\,A* and size measurements at 43\,GHz}
Here we show preliminary results from ongoing analysis. As shown in Figure~1 (right-hand panel), all the images show emission east-west oriented along a PA of $\sim$ 80$^\circ$. We find no evidence for a significant deviation of the closure phase from zero within $\pm10^\circ$ at the three frequencies. The 43\,GHz data are very well fitted by a single elliptical Gaussian component. These models are given in Table~2. The apparent source size at 43\,GHz (and thus the intrinsic size for an assumed stable scattering model, no variations of the ISM) appears to be constant on a time scale of 10\,days and remains even unchanged after a decade of observation, when compared to the data of the last 15\,years (\cite{kri93}, \cite{lo}, \cite{bower04}, \cite{shen}) despite a refractive timescale of the scattering medium of less than 1\,year. From the average overall observing epochs, we obtained the following mean source parameters: $\theta_{major}$ = 0.71 $\pm$ 0.01 mas, Ratio = 0.55 $\pm$ 0.01, and PA = 79.6 $\pm$ 0.3$^\circ$.
\begin{figure}[t]
\centering
\begin{minipage}{28pc}
\includegraphics[width=28pc]{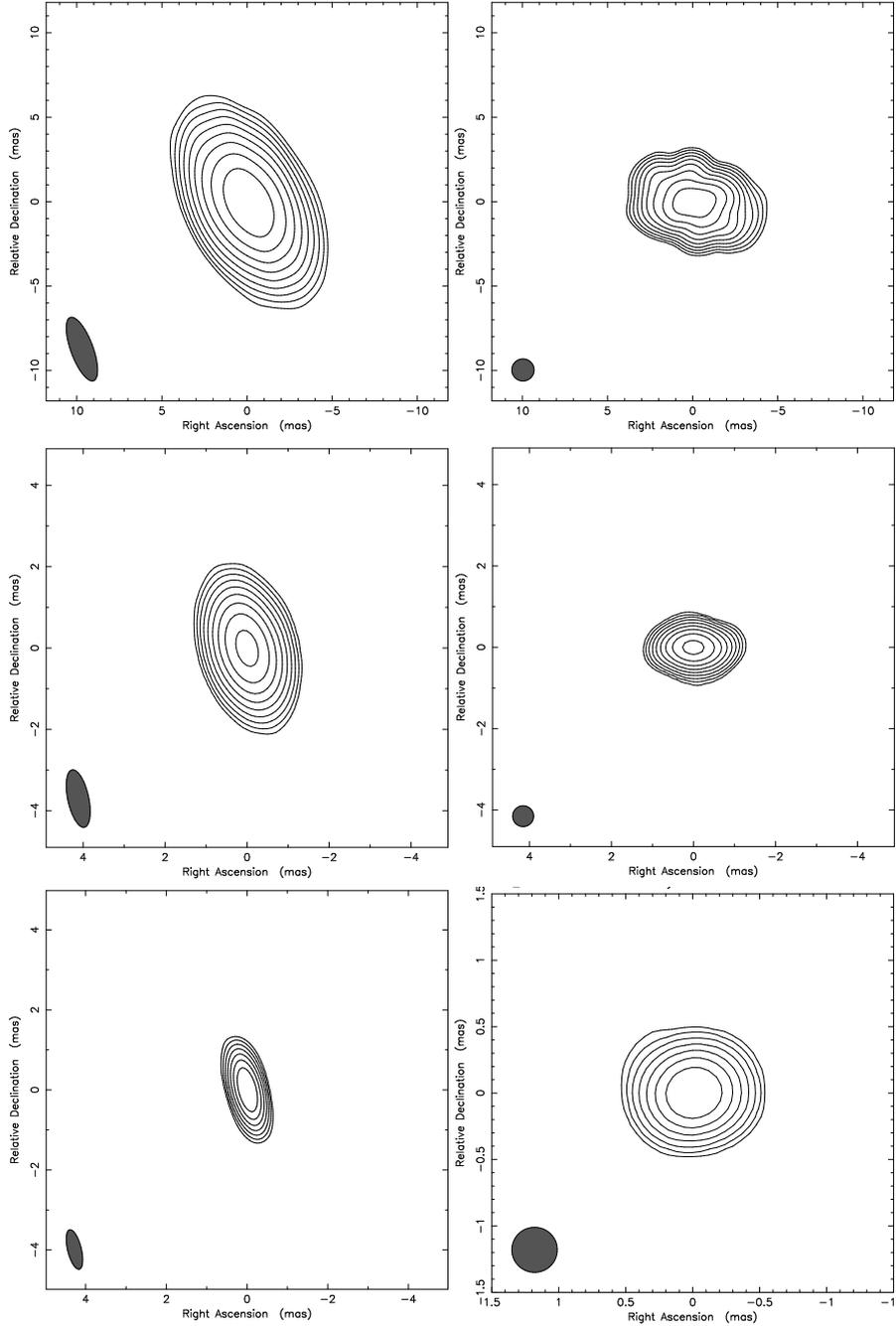}
%\caption{\label{3mm}Figure caption for first of two sided figures.}
\end{minipage}
\caption{Uniformly weighted VLBA images of Sgr\,A* on May 15, 2007  at 22 (top panels), 43 (middle panels) and 86\,GHz (bottom panels), respectively. At each row of panels, the right-hand panel shows a slightly super-resolved image restored with a circular beam corresponding to the minor axis of the elliptical beam of the left-hand image. The contour levels are the same as that in the left-hand panel. The parameters of the images are listed in Table~1.}
\end{figure}

%\clearpage
\begin{table}[t]
\caption{\label{jfonts}Description of VLBA images of Sgr\,A* shown in Figure~1.} 
\begin{center}
\lineup
\begin{tabular}{@{}c*{15}{c}}
\br
&&\multicolumn{3}{c}{Restoring Beam}\\
\cline{3-5}
Frequency&S$_{peak}$&Major&Minor&PA&Contours\\
\mbox{[GHz]}&[Jy/beam]&[mas]&[mas]&[deg]&\\
(1)&(2)&(3)&(4)&(5)&(6)\\
\mr
22&0.653 (0.456)&4&1.31&20.1&0.2, 0.4, 0.8, 1.6, 3.2, 6.4, 12.8, 25.6, 51.2\\
43&1.20 (1.08)&1.44&0.513&12.1&0.2, 0.4, 0.8, 1.6, 3.2, 6.4, 12.8, 25.6, 51.2\\
86&2.98 (2.89)&1.01&0.34&13.7&0.2, 0.4, 0.8, 1.6, 3.2, 6.4, 12.8, 25.6, 51.2\\
\br
\end{tabular}
\end{center}
\scriptsize {Notes: (1) Observing frequency; (2) Peak flux density, the numbers in brackets correspond to the peak intensity of images in the right panel of each row in Figure~1;
(3), (4), (5) Parameters of the restoring elliptical Gaussian beam: the full width at half maximum (FWHM) of the major and minor axes and the position angle (PA) of the major axis.
(6) Contour levels of the image, expressed as a percentage of the peak intensity.}
\end{table}

\begin{table}[h]
\caption{\label{size} Results from the model fitting of the 43\,GHz data for the ten epochs of observations.}
\begin{center}
\lineup
\begin{tabular}{*{7}{c}}
\br
Date&Flux&Major&Ratio&PA\cr
[\mbox{May 2007}] & [Jy]&[mas]&&[deg]\cr
\mr
\0\015&2.02$\pm$0.09&0.71$\pm$0.01&0.58$\pm$0.06&81.7$\pm$2.3 \cr
\0\016&1.59$\pm$0.07&0.72$\pm$0.01&0.53$\pm$0.01&82.1$\pm$0.8 \cr
\0\017&1.99$\pm$0.06&0.72$\pm$0.01&0.62$\pm$0.05&82.0$\pm$2.4 \cr
\0\018&1.61$\pm$0.04&0.71$\pm$0.01&0.54$\pm$0.09&84.1$\pm$1.1 \cr
\0\019&1.86$\pm$0.07&0.71$\pm$0.01&0.51$\pm$0.06&84.7$\pm$3.1 \cr
\0\020&1.66$\pm$0.06&0.72$\pm$0.01&0.54$\pm$0.06&80.4$\pm$1.6 \cr
\0\021&2.02$\pm$0.08&0.72$\pm$0.01&0.62$\pm$0.06&81.8$\pm$3.1 \cr
\0\022&1.90$\pm$0.05&0.72$\pm$0.01&0.66$\pm$0.06&78.5$\pm$0.3 \cr
\0\023&1.92$\pm$0.08&0.72$\pm$0.01&0.54$\pm$0.06&81.2$\pm$1.3 \cr
\0\024&1.78$\pm$0.06&0.68$\pm$0.01&0.48$\pm$0.06&86.6$\pm$2.1 \cr
\mr
\0\0weighted&mean: &0.71$\pm$0.01&0.55$\pm$0.01&79.6$\pm$0.3 \cr
\br
\end{tabular}
\end{center}
\scriptsize {Notes: Listed are the observing date in May 2007, total flux density in [Jy], major axis of the elliptical Gaussian in [mas], the ratio of the minor axis to the major axis, and the position angle of the major axis.}
\end{table}

\section{Flux variations at 43\,GHz}
As shown in Figure~2, we plot the light curve of Sgr\,A* at 43\,GHz on a daily basis.
With the modeled flux densities from both right-hand (RR) and left-hand (LL) correlations, we derived small flux correction factors for Sgr\,A* assuming stationarity of the flux of our secondary calibrator NRAO\,530 over 10 days. Figure~2 (left) shows the uncorrected VLBI flux densities of Sgr\,A* at each epoch, and on the right the corresponding corrected values. Clearly, Sgr\,A* has undergone a flux density variation with reduced $\chi_\nu^2$ of 8.26 and 6.80 for the flux densities from RR and LL correlations, which gives probabilities of 1.1 $\times$ 10$^{-22}$ and 5.6 $\times$ 10$^{-20}$ for the source being non-variable. The light curve indicate that flux variations are more pronounced at the beginning of the campaign, which just happened to coincide with two NIR flares on 15th and 17th of May (see also Kunneriath \textit{et al}, this conference). We see that for both sources the LL flux is slightly larger than the RR flux, indicating a small uncorrected instrumental polarization offset, consistent with zero circular polarization of either source.
\begin{figure}[t]
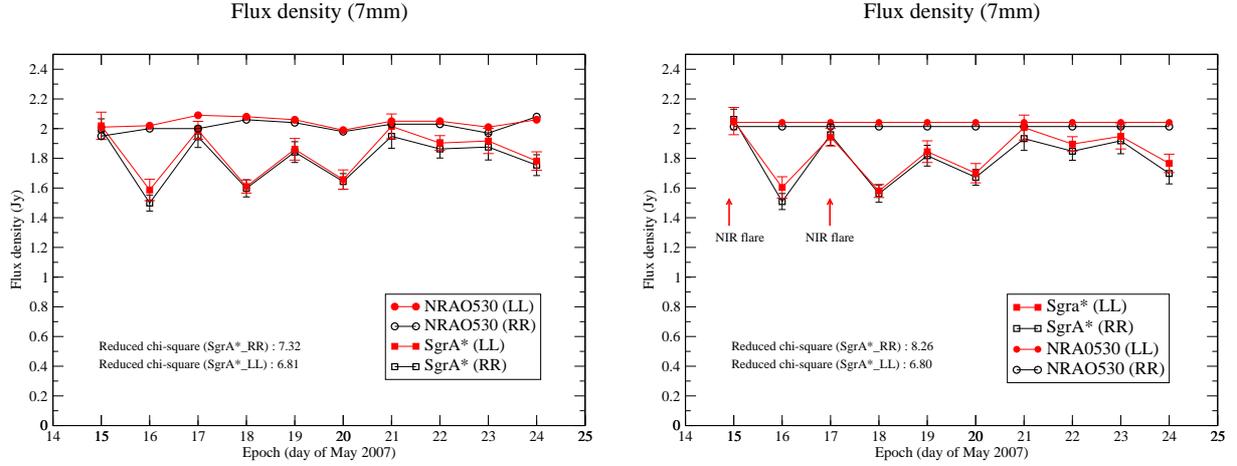

\centering
%\begin{minipage}{16pc}
\includegraphics[clip,width=0.48\textwidth]{lu_fig2.eps}
%\end{minipage}\hspace{2pc}%
%\begin{minipage}{16pc}
\hfill
\includegraphics[clip,width=0.48\textwidth]{lu_fig3.eps}
%\end{minipage}
\caption{\label{flux} Left: 43\,GHz VLBI flux variations of Sgr\,A* (squares). The flux density of NRAO\,530 (circles) is shown for comparison. The empty symbols represent the flux densities from RR and filled symbols are the flux densities from LL. Right: Same plot but with flux densities of Sgr\,A* corrected by small factors derived from the normalization of the flux density of NRAO\,530 to its mean value.}
\end{figure}

\section{Spectrum and structure of Sgr\,A*}
In Figure~3 (left), we plot the averaged VLBI spectrum of Sgr\,A* from the available published data in the 1990s, and the spectrum from the new measurements at the first epoch. The source got $\sim$ 2 times brighter roughly maintaining the same spectral index. With respect to the constant source size (see below), this also indicates an increase of the brightness temperature by a factor of $\sim$  2.

Shown in Figure~3 (right) is the apparent source size (FWHM) of Sgr\,A* at 22, 43, and 86\,GHz of the major and minor axis, respectively. At 22 and 86\,GHz, we only use the first epoch data, while at 43\,GHz, the mean value is used. The slope for the minor axis deviates from the scattering law towards the shorter wavelengths, as shown in Figure~3 (right, dashed line). This may indicate that some intrinsic east-west oriented structure shines through. At this moment, the measured apparent sizes are consistent with the known $\lambda^2$ dependence from scattering. Assuming validity of the scattering law of \cite{bower06}, we tentatively extract in quadrature the intrinsic major source size of 87, 32, and 14 Schwarzschild radii at 22, 43, and 86\,GHz respectively (assuming a 4 $\times$ 10$^6$ solar mass BH). These numbers agree well with values recently reported by \cite{kri06} and \cite{yuan}.

\begin{figure}[t]
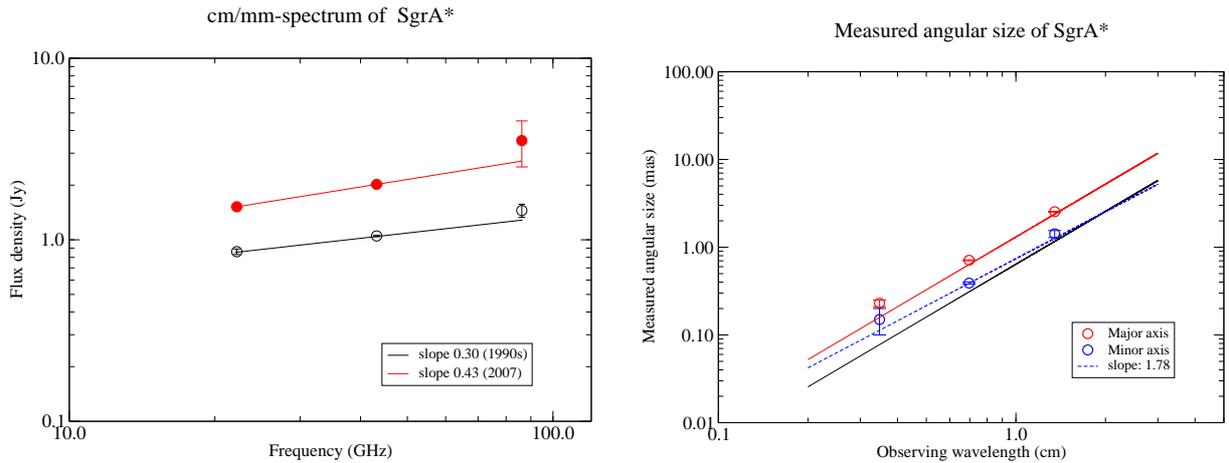

%\centering
%\begin{minipage}{16pc}
\includegraphics[width=0.48\textwidth]{lu_fig4.eps}
%\end{minipage}
%\hspace{2pc}%
%\begin{minipage}{16pc}
\hfill
\includegraphics[width=0.48\textwidth]{lu_fig5.eps}
%\end{minipage}
\caption{\label{size} Left: The 3 frequency radio spectrum of Sgr\,A*. Open circles show an averaged mean radio spectrum from VLBI observations in the 1990s. Filled circles denote the fluxes from our new measurements on May 15 2007 (first epoch). Solid lines show the power-law fits to the VLBI spectra. Right: Measured angular size of Sgr\,A* plotted versus wavelength. The solid lines delineates the scattering model for the major and minor axis of a 1.309 $\lambda^2$ and 0.64 $\lambda^2$ law \cite{bower06}. The dashed line represents a power-law fit of index 1.78 to the extension of the minor axis measured at the 3 frequencies.}
\end{figure}

\section{Summary}
We have shown images and parameters from 10 epochs of high frequency VLBI observations of Sgr\,A* during a multi-frequency campaign in May 2007. In particular, we have measured the flux density variations and shown the size measurement at 43\,GHz. Our result indicate that the measured size at 43\,GHz is stable on a daily basis. Future studies will show if there is a correlation between the NIR flaring activities and the radio flux density variations. The radio (cm-mm-wavelengths) flux of Sgr\,A* is known to be variable on timescales of months to hours, with variability being more pronounced at shorter wavelengths. The time scales of the variations indicate that the higher frequency emission originates from a very compact region (e.g.,~\cite{Miya}, ~\cite{Mau}). The mm-variability naturally raises the question of whether or not the flux density variations are accompanied by the structural change, which could be detected with VLBI at 86\,GHz. Previous size measurements at 86\,GHz may have already indicated possible structural variability, where the increase in size seems to correlate with a higher flux density level~\cite{kri06}. However, these variations of the source size is at present uncertain and still requires confirmation. As such, the 86\,GHz VLBI measurements from this campaign will be very crucial for the possible structural variability.

\ack %Based on observations with the 100-m radio telescope of the MPIfR (Max-Planck-Institut für Radioastronomie) at Effelsberg.%
R-S Lu and D Kunneriath are members of the International Max Planck Research School (IMPRS) for Astronomy and Astrophysics. The National Radio Astronomy Observatory is a facility of the National Science Foundation operated under cooperative agreement by Associated Universities, Inc.

\end{document}